\documentstyle[sprocl,epsfig]{article}

\def\be{\begin{equation}}
\def\ee{\end{equation}}
\def\bea{\begin{eqnarray}}
\def\eea{\end{eqnarray}}

\begin{document}

\title{ELECTROMAGNETIC PROBES OF DENSE MATTER IN 
RELATIVISTIC HEAVY-ION COLLISIONS\footnote{Based on invited talk 
presented at the APCTP Workshop on Hadron Properties in Medium,
in Honor of Manque Rho's 60th Birthday, Seoul, October 27-31, 1997.}}

\author{G.Q. Li, G.E. Brown, C. Gale\footnote{Permanent address: Physics 
Department, McGill University, Montreal, QC, H3A 2T8, Canada}}
\address{Department of Physics and Astronomy,  
        State University of New York\\
       at Stony Brook, Stony Brook, NY 11793, USA\\
       Email: gqli@nuclear.physics.sunysb.edu\\
       Email: popenoe@nuclear.physics.sunysb.edu\\
       Email: gale@hendrix.physics.mcgill.ca}
\author{C.M. Ko}
\address{Cyclotron Institute and Department of Physics, Texas
A\&M University,\\
College Station, TX 77843, USA\\
Email: ko@comp.tamu.edu}

\maketitle\abstracts{
Dilepton and photon production in heavy-ion collisions at 
SPS energies is studied in the relativistic transport 
model that incorporates self-consistently the change of hadron
masses in dense matter. It is found that the dilepton spectra in 
proton-nucleus reactions can be well described by the conventional
mechanisms of Dalitz decay, primary vector meson decay, decay of
charmed mesons, and the initial Drell-Yan processes. However, 
to provide a quantitative explanation of the observed dilepton spectra 
in central heavy-ion collisions requires contributions other than 
these direct decays and also various medium effects. 
Introducing a decrease of vector meson masses in hot dense medium, we find 
that the low-mass dilepton enhancement can be satisfactorily explained. 
Furthermore, to explain the intermediate-mass dilepton enhancement
in heavy-ion collisions, secondary processes such as $\pi a_1\rightarrow
l{\bar l}$ are found to be very important.
Finally, the single photon spectra in our calculations with 
either free or in-medium meson masses do not exceed the upper 
limit measured by the WA80 Collaboration.
}

\section{INTRODUCTION}
 
The experimental measurement and theoretical investigation of
electromagnetic observables in heavy-ion collisions constitutes one
of the most active and exciting fields in physics \cite{qm96}.
Because of their relatively weak final-state interactions
with the hadronic environment, dileptons, as well as photons, are
considered ideal probes of the early stage of heavy-ion
collisions, where quark-gluon-plasma (QGP) formation and
chiral symmetry restoration are expected \cite{shur80,kkmm86,brown96}.  

Dilepton mass spectra can be basically divided into three mass regions.
The low-mass region below $m_\phi$ ($\sim$ 1 GeV)
is dominated by hadronic interaction and hadronic
decay in the freeze-out. In the intermediate-mass region
between $m_\phi$ and about 2.5 GeV, the contribution
from the thermalized QGP might be seen. In the high-mass region at
and above $m_{J/\Psi}$ the major effort has been the
detection and understanding of $J/\Psi$ suppression.

So far, the experimental measurement of dilepton spectra 
in heavy-ion collisions has mainly been carried out
at the CERN SPS by three collaborations: the CERES
collaboration is dedicated to dielectron spectra in the 
low-mass region \cite{ceres,drees96}, the HELIOS-3 \cite{helios} 
collaboration has measured dimuon spectra from the its 
threshold up to the $J/\Psi$ region, and the NA38/NA50 \cite{na38} 
collaboration measures dimuon spectra in the intermediate- 
and high-mass region, emphasizing the $J/\Psi$ supression.
 
Recent observation of the enhancement of low-mass dileptons in
central heavy-ion collisions by the CERES \cite{ceres,drees96} and
the HELIOS-3 \cite{helios} collaborations has generated a 
great deal of interest in the heavy-ion community. 
Different dynamical models have been used to calculate
and understand these data. The results from many groups with 
standard scenarios (i.e., using vacuum meson properties) are 
in remarkable agreement with each other, but in significant 
disagreement with the data: the experimental spectra in the 
mass region from 0.3-0.6 GeV are substantially underestimated 
\cite{likob,cassing,rapp,sri96,soll96,hung97,sch96,other} 
(see also Ref. \cite{drees96}). This has led to the suggestion of 
various medium effects that might be responsible for the
observed enhancement. In particular, the dropping vector 
meson mass scenario \cite{likob,br91,hat92} due to 
the chiral symmetry restoration is found to provide a nice 
description of both the CERES and HELIOS-3 data. On the
other hand, conventional many-body effects leading to a brodening
of the rho-meson spectrum function also provide a good
description of these data \cite{rapp}. 

In the high-mass region around $m_{J/\Psi}$, the $J/\Psi$ suppression 
has been a subject of great interest, since it was first proposed
as a signal of the deconfinement phase transition \cite{satz86}. Various
investigations show that up to central S+Au collisions, the normal
pre-resonance absorption in nuclear matter is sufficient to account 
for the observed $J/\Psi$ suppression. However,
recent data from the NA50 collaboration for central Pb+Pb
collisions show an additional strong `anomalous' suppression
which might indicate the onset of the color deconfinement
\cite{na50}. 

Enhancement of dileptons in the intermediate-mass region from 
about 1 GeV to about 2.5 GeV has also been observed by 
both the HELIOS-3 and NA38/NA50 collaborations
in central S+W and S+U collisions as compared to that in the 
proton-induced reactions (normalized by the charged-particle 
multiplicity) \cite{helios,na38}. Preliminary data from the
NA50 collaboration also show significant enhancement in central
Pb+Pb collisions \cite{na38} (see also Ref. \cite{drees96}).
The intermediate-mass dilepton spectra in heavy-ion collisions
are particularly useful for the search of the QGP. It was 
originally suggested that in the intermediate-mass
or the intermediate $p_t$ region,
the electromagnetic radiation from the QGP phase might shine
over that from the hadronic phase \cite{shur78,kap91}.
However, to extract from the measured dilepton spectra any 
information about the phase transition and the properties of
the QGP, it is essential that the contributions from the
hadronic phase be precisely understood and carefully
subtracted.

Another piece of experimental data from CERN SPS that has been 
discussed extensively is the single photon spectra from the WA80 
collaboration \cite{wa80}. In this experiment, $\pi^0$ and $\eta$
spectra were measured simultaneously, so that their contributions
to the single photon spectra can be subtracted. 
It was found that the direct photon excess over those background sources 
is about 5\% for central S+Au collisions \cite{wa80}.
Similar measurement of inclusive photon spectra has been 
carried out by the CERES collaboration \cite{baur96}.
The results are very much in agreement with those of WA80, 
namely, the inclusive photon spectra can be basically 
explained by hadronic decay at the freeze-out, particular 
radiative decay of $\pi^0$ and $\eta$.

In various hydrodynamics calculations 
the absence of significant thermal photon production has been used as 
an indication of quark gluon plasma formation that lowers the 
initial temperature \cite{sinha94,dumi95}. However, there is a
major uncertainty in this type of analysis, namely, the initial
temperature depends sensitively on the degrees of freedom 
included in the analysis. Indeed, in Refs. \cite{soll96,cley97},
it was found that including a sufficient amount of hadron 
resonances the WA80 photon data can be explained without
invoking the formation of the QGP. Detailed transport model 
calculations \cite{lib97} support the findings of 
Refs. \cite{soll96,cley97}.

In this contribution, we will discuss (1) the enhancement of
low-mass dileptons (Section 2), (2) the enhancement of intermediate-mass
dileptons (Section 3), and (3) the lack of direct photon signal
(Section 4). A brief summary and outlook is presented
in Section 5.  

\section{Low-mass dilepton enhancement: a case for dropping rho meson
mass}

To calculate dilepton spectra in heavy-ion collisions and to
studying possible medium effects, the relativistic 
transport model \cite{koli96} based on the Walecka-type model \cite{qhd86}
has been quite useful, as it provides a thermodynamically consistent 
description of the medium effects through the scalar and vector
fields. In heavy-ion collisions at SPS energies, 
many hadrons are produced in the initial nucleon-nucleon 
interactions. This is usually modeled by the fragmentation of 
strings, which are the chromoelectric flux-tubes 
excited from the interacting quarks. One successful model 
for taking into account this nonequilibrium dynamics is the RQMD model 
\cite{sorge89}. To extend the relativistic transport model to heavy-ion 
collisions at these energies, we have used as initial conditions the 
hadron abundance and distributions obtained from the string fragmentation 
in RQMD.  Further interactions and decays of these hadrons are then 
taken into account as in usual relativistic transport model.

To study the effects of dropping vector meson masses 
\cite{br91,hat92} on the dilepton spectrum in heavy-ion 
collisions, we have extended the Walecka model from the 
coupling of nucleons to scalar and vector fields to the 
coupling of light quarks to these fields, using the 
ideas of the meson-quark coupling model \cite{thomas94}.
For a system of nucleons, pseudoscalar mesons,
vector mesons, and axial-vector mesons
at temperature $T$ and baryon density $\rho _B$, the scalar field 
$\langle\sigma\rangle$ is determined self-consistently from 
\begin{eqnarray}
m_\sigma^2\langle \sigma\rangle &=&{4g_\sigma\over (2\pi )^3}\int d{\bf k} 
{m_N^*\over E^*_N}\Big[{1\over \exp ((E^*_N-\mu _B)/T)+1}
+{1\over \exp ((E^*_N+\mu _B)/T)+1}\Big]\nonumber\\
&+&{0.45g_\sigma\over (2\pi )^3}\int d{\bf k} {m_\eta^*\over E_\eta ^*}
{1\over \exp (E_\eta ^*/T)-1}+
{6g_\sigma\over (2\pi )^3}\int d{\bf k} {m_\rho^*\over E_\rho ^*}
{1\over \exp (E_\rho ^*/T)-1}\nonumber\\
&+&{2g_\sigma\over (2\pi )^3}\int d{\bf k} {m_\omega^*\over E_\omega ^*}
{1\over \exp (E_\omega ^*/T)-1}
+{6\sqrt 2 g_\sigma\over (2\pi )^3}\int d{\bf k} {m_{a_1}^*\over E_{a_1}^*}
{1\over \exp (E_{a_1}^*/T)-1},
\end{eqnarray}
where we have used the constituent quark model relations for 
the nucleon and vector meson masses, 
i.e., $m_N^*=m_N-g_\sigma\langle \sigma\rangle ,
~m_{\rho ,\omega}^*\approx m_{\rho ,\omega}
-(2/3)g_\sigma\langle\sigma\rangle$, the quark 
structure of the $\eta$ meson in free space which leads to 
$m_\eta^*\approx m_\eta -0.45g_\sigma\langle\sigma\rangle$, and 
the Weinberg sum rule relation between the rho-meson and $a_1$ 
meson masses, 
i.e, $m_{a_1}^*\approx m_{a_1}-(2\sqrt 2/3)g_\sigma\langle\sigma\rangle$.

The main contributions to dileptons with mass below 1.2 GeV are the
Dalitz decay of $\pi^0$, $\eta$ and $\omega$, the direct leptonic decay 
of the primary $\rho^0$, $\omega$ and $\phi$, the pion-pion annihilation 
which proceeds through the $\rho^0$ meson, and the kaon-antikaon 
annihilation that proceeds through the $\phi$ meson. The last two 
processes are unique in heavy-ion collisions, and especially the 
pion-pion annihilation is found to be very important for low-mass 
dilepton production. The explicit treatment of rho and phi meson
formation, propagation, and decay in pion-pion and kaon-antikaon
annihilation avoids possible double counting.

The differential widths for the Dalitz decay of $\pi^0$, $\eta$, 
and $\omega$ are related to their radiative decay widths via 
the vector dominance model, which are taken from Ref. \cite{land85}.  
The decay of a vector meson into the dilepton is determined by the width,
\begin{equation}\label{lwidth}
\Gamma_{V\to l^+l^-}(M)= C_{l^+l^-}
\frac{m_V^4}{3M^3}(1-\frac{4m_l^2}{M^2})^{1/2}(1+\frac{2m_l^2}{M^2}).
\end{equation}
The coefficient 
$C_{l^+l^-}$ in the dielectron channel is $8.814\times 10^{-6}$,
$0.767\times 10^{-6}$, and $1.344\times 10^{-6}$ for $\rho$, $\omega$,
and $\phi$, respectively, and is determined from the measured width.
For the dimuon channel, these values are slightly larger.
The two-body processes (pion-pion and kaon-antikaon annihilation)
are treated in two steps: the formation of the intermediate vector
meson and its decay into dileptons.

\begin{figure}[htb]
\begin{center}
\epsfig{file=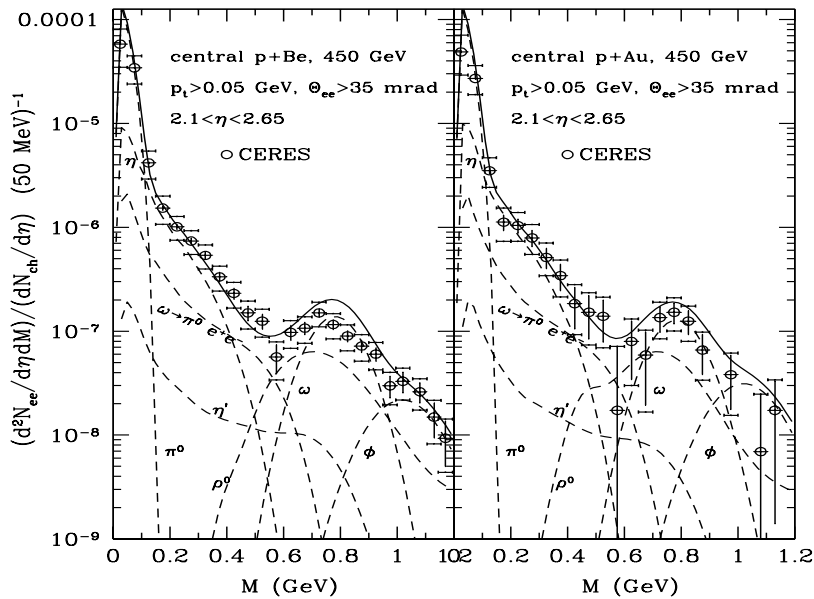,height=3.5in,width=4.5in}
\caption{Dilepton invariant mass spectra in p+Be (left window)
and p+Au (right window) collisions at 450 GeV.}
\end{center}
\end{figure}

The results for dilepton spectra from p+Be and p+Au collisions 
at 450 GeV are shown in Fig. 1, together with data from the 
CERES \cite{ceres}. It is seen that the data can be well 
reproduced by Dalitz decay of $\pi^0$, $\eta$ and $\omega$ 
mesons, and direct leptonic decay of $\rho^0$, $\omega$ 
and $\phi$ mesons. These results are thus similar 
to those found in Ref. \cite{cassing} using the Hadron-String Dynamics
and those constructed by the CERES collaboration from known and expected 
sources of dileptons \cite{ceres}. 

\begin{figure}[htb]
\begin{center}
\epsfig{file=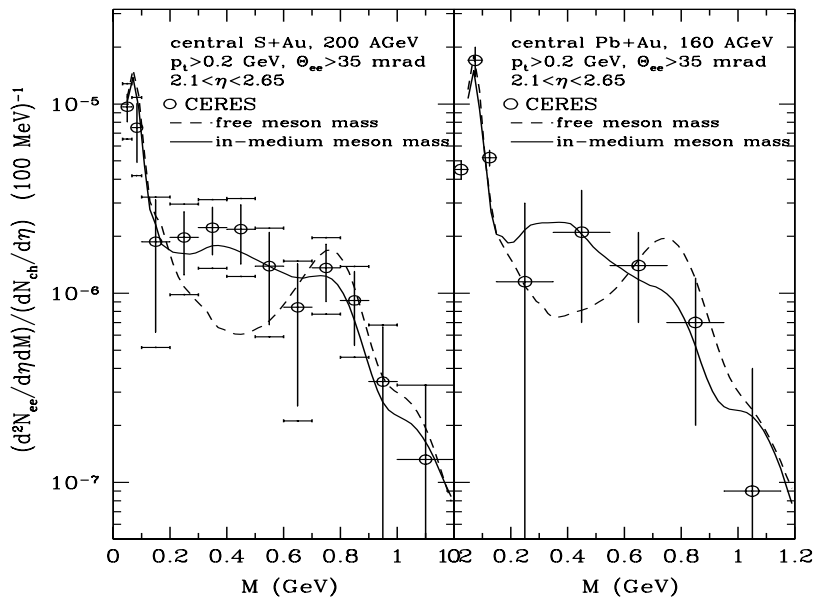,height=3.5in,width=4.5in}
\caption{Dilepton invariant mass spectra in S+Au collision at 
200 AGeV (left window) and Pb+Au collisions at 160 AGeV 
(right window). The solid and dashed curves are obtained with
in-medium and free meson masses, respectively.}
\end{center}
\end{figure}

Our results for dilepton spectra in central S+Au collisions
are shown in the left window of Fig. 2. The dashed 
curve is obtained with free meson masses. Although 
pion-pion annihilation is important for dileptons with 
invariant mass from 0.3 to 0.65 GeV, it still does not 
give enough number of dileptons in this mass region.
Furthermore, for masses around $m_{\rho,\omega}$ there are 
more dileptons predicted by the theoretical calculations than shown in 
the experimental data. These are very similar to the results 
of Cassing {\it et al.} \cite{cassing} based on the 
Hadron-String Dynamics model and Srivastava {\it et al.} \cite{sri96}
based on the hydrodynamical model.
The results obtained with in-medium meson masses are shown 
by the solid curve.  Compared with the results obtained with free meson 
masses, there is about a factor of 2-3 enhancement of the dilepton yield 
in the mass region from 0.2 to 0.6 GeV, which thus leads to a good 
agreement between the theoretical results and the CERES data. 
Similar conclusions that dropping vector
meson masses can explain the CERES dilepton data have been
obtained in Refs. \cite{cassing,hung97,sch96}. 
 
In the right window of Fig. 2 we compare our prediction for
central Au+Pb collisions at 160 AGeV with the preliminary
data from the CERES collaboration. 
The normalization factor $dN_{ch}/d\eta$ here is the average charge 
particle pseudo-rapidity density in the pseudo-rapidity range of 2 
to 3, and is about 440 in this collision. In the results with free meson 
masses, shown by the dashed curve, there 
is a strong peak around $m_{\rho ,\omega}$, which is dominated by 
$\rho^0$ meson decay as a result of an enhanced contribution from 
pion-pion annihilation in Pb+Au collisions than in S+Au and 
proton-nucleus collisions. In the case of in-medium meson masses, 
shown by the solid curve, the $\rho$ 
meson peak shifts to a lower mass, and the peak around $m_{\rho ,\omega}$ 
becomes a shoulder arising mainly from $\omega$ meson decay. At the same 
time we see an enhancement of low-mass dileptons in the region of 
0.25-0.6 GeV as in S+Au collisions. The agreement with the data
is thus significantly improved when dropping vector meson masses are used.      

\begin{figure}[htb]
\begin{center}
\epsfig{file=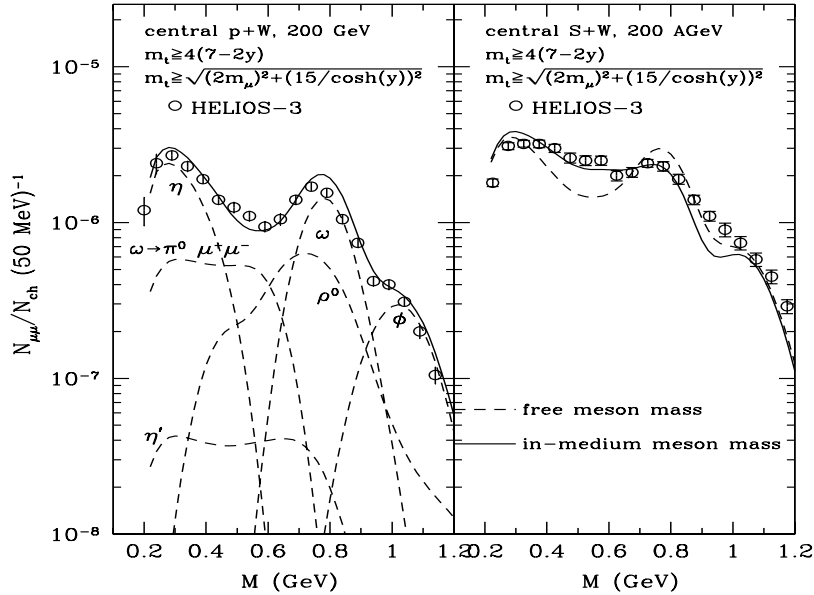,height=3.5in,width=4.5in}
\caption{Dilepton invariant mass spectra in p+W  
(left window) and S+W (right window) collisions at 200 AGeV. 
In the right window the solid and dashed curves are obtained with
in-medium and free meson masses, respectively.}
\end{center}
\end{figure}

The same model has been used to calculate the dimuon spectra 
from p+W and central S+W collisions. The results for 
p+W are shown in the left window of Fig. 3. Again, the
conventional mechanism describes quite well the 
measured spectra. The results for S+W collisions obtained
with free meson masses are shown in the right window of Fig. 3
by the dashed curve, and are below the HELIOS-3 data in the mass 
region from 0.35 to 0.6 GeV, and slightly above the data 
around $m_{\rho ,\omega}$ as in the CERES case. However, 
the discrepancy between the theory and the data is somewhat 
smaller in this case due to the smaller charged-particle multiplicity at 
a larger rapidity than in the CERES experiment.
Our results obtained with in-medium meson masses are shown in 
the right window of Fig. 3 by the solid curve, and are 
in good agreement with the data.  The importance of 
dropping rho meson mass in explaining the HELIOS-3 data has also been 
found by Cassing {\it et al.} \cite{cassing}.

It should be pointed out that, the effects of dropping rho meson
mass on the low-mass dilepton spectra, as seen in Figs. 2 and 3,
were predicted theoretically before the CERES data were published.
In Ref. \cite{rqmddi}, the RQMD model was used to calculate
dilepton spectra in Au+Au collisions at AGS energies. 
It was found that, with an in-medium rho meson mass as 
predicted by the QCD sum rule calculation, the rho meson
peak in the dilepton spectra shifts to about 0.5 GeV, and
a significant enhancement is seen in this mass region.
Similar observation was obtained in Ref. \cite{liko95} 
for dilepton spectra in heavy-ion collisions at SIS
energies.
 
Rho meson decays dominantly into two pions. Its spectral
function can therefore be modified by the medium modification
of the pion dispersion function. Furthermore, rho meson
itself can interact with nucleons to form nucleon resonances,
in particular $N(1720)$ and $\Delta (1905)$, which have appreciable
decay branching ratios into the $\rho N$ final state.
Both effects lead to the broadening of the rho meson spectral
function and thus the enhancement of low-mass dileptons
from the in-medium rho meson decay. This idea was shown
in Ref. \cite{rapp} to also lead to a reasonable description of
the low-mass dilepton enhancement observed by the CERES
collaboration. 
 
\section{Intermediate-mass dileptons: the importance of secondary 
processes}

To extend the previous calculation to the intermediate mass
region, we need to include additional secondary processes
in addition to $\pi\pi$ and $K{\bar K}$ annihilation,
as well as the contributions from the charmed mesons and
initial Drell-Yan processes. The last two contributions
involve hard processes which
scale almost linearly with the participant nucleon number,
and can thus be extrapolated from the proton-proton and
proton-nucleus collisions. Such a study has recently been
carried out by Braun-Munzinger {\it et al} \cite{braun97}.
The results for the central S+W collisions corresponding to
the HELIOS-3 acceptance are shown in the left window of Fig. 4,
which are taken from Ref. \cite{drees96}. These, together
with the dileptons from the decays of primary vector mesons,
are collectively labeled `background'. It is seen that these
background sources describe very well the dimuon spectra in the
p+W reactions, shown in the figure by solid circles.

However, as can be from the figure, the sum of these background
sources grossly underestimates the dimuon yield in central
S+W collisions, shown in the figure by open circles.
Since the dimuon spectra are normalized by the measured
charged particle multiplicity, this underestimation indicates
additional sources to dilepton production in heavy-ion
collisions. This can come from the thermalized QGP and/or
hadronic phases. So the immediate next step is to check 
whether the contribution from the secondary hadronic processes
can explain this enhancement. For dilepton spectra
at low invariant masses, it is well known that the
$\pi\pi$ annihilation plays an extremely important role in
heavy-ion collisions. It is also expected that the
other secondary processes will play a role in the
dilepton spectra in the intermediate mass region. 

Previous thermal rate calculations based on the kinetic theory
show that in the mass and temperature region relevant for
this study, the following secondary processes (from the hadronic
phase) are very important: $\pi\pi\rightarrow l{\bar l}$,
$\pi\rho\rightarrow l{\bar l}$, $\pi\omega\rightarrow l{\bar l}$, 
$\pi a_1\rightarrow l{\bar l}$, 
$K{\bar K}\rightarrow l{\bar l}$, and $K{\bar K^*}+c.c \rightarrow 
l{\bar l}$ \cite{gale94,song94,haglin95,kim96}. 
Among them, the $\pi a_1\rightarrow l{\bar l}$ has 
been found to be the most important, mainly because of its
large cross section \cite{song94,kim96} (Similar conclusion
has been drawn for thermal photon production \cite{xiong92,song93}).

\begin{figure}[htb]
\begin{center}
\epsfig{file=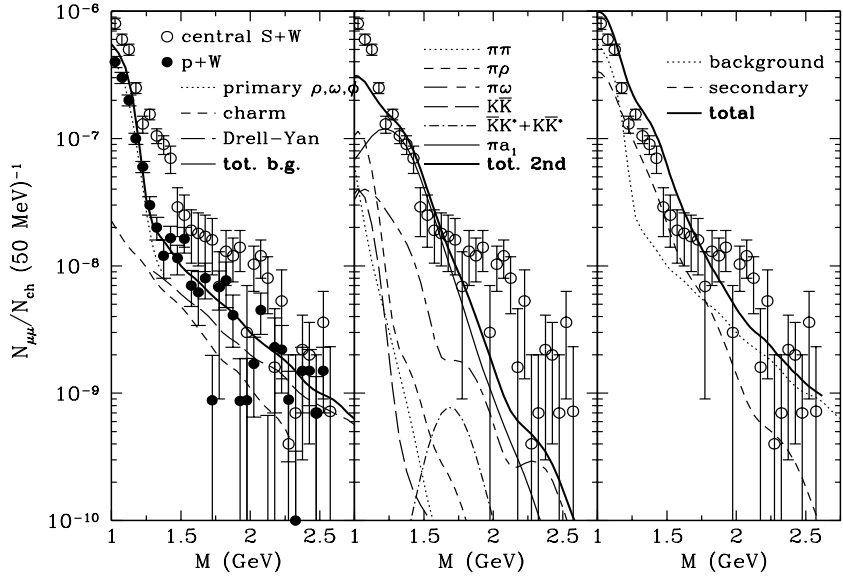,height=5in,width=3.5in,angle=270}
\caption{Left panel: comparison of backgrouds with experimental data
in p+W and S+W collisions. Middle panel: contributions of various
secondary processes to the dimuon spectra in central S+W
collisions. Right panel: Comparison of the sum of the background
and secondary contributions to the experimental data in central
S+W collisions.}
\end{center}
\end{figure}

We will concentrate on dilepton production from the secondary
processes mentioned above. The cross sections for the annihilation
of pseudoscalar mesons ($P$) are well known
\begin{eqnarray}
\sigma _{PP\rightarrow l{\bar l}} (M)
={8\pi \alpha^2 k \over 3M^3} |F_P(M)|^2 (1-{4m_l^2\over M^2})
(1+{2m_l^2\over M^2}),
\end{eqnarray}
where $k$ is the magnitude of the three-momentum of the pseudoscalar
meson in the center-of-mass frame, $M$ is the mass of the
dilepton pair, and $m_l$ is the mass of the lepton. It is well
known that the electromagnetic form factors $|F_P(M)|^2$
play important role in these processes. The pion electromagentic
form factor is donimated by the $\rho (770)$ meson, 
while that of the kaon is dominated by the phi meson. 
In addition, at large invariant masses, higher $\rho$-like 
resonances such as $\rho (1450)$ were found to be important \cite{bia91}.

The cross sections for $\pi\rho\rightarrow l{\bar l}$ and 
${\bar K}K^*+c.c.\rightarrow l{\bar l}$, which are of the
pseudoscalar-vector type, have been studied in Ref. \cite{haglin95}
by fitting to the experimental data for the reverse processes
(i.e., hadron production in $e^+e^-$ annihilation) \cite{donn89,mane82}.
High isoscalar vector mesons such as $\omega (1420)$ and $\phi (1680)$
play important roles in the electromagnetic form factors
of these processes. The cross sections from Ref. \cite{haglin95}
will be used in this work.
The cross section for $\pi\omega \rightarrow l{\bar l}$ has the same 
form as that for $\pi\rho\rightarrow l{\bar l}$, but with a
different form factor, which is
determined from the experimental data for $e^+e^-
\rightarrow \pi^0\pi^0\gamma$, as shown in Ref. \cite{doli91}.

The cross section for $\pi a_1\rightarrow l{\bar l}$ needs some 
special attention, since this process has been found to be
particularly important in the intermediate-mass region.
Already in Ref. \cite{kim96} it was shown that the
dilepton production rates based on different models
for the $\pi\rho a_1$ dynamics can differ by an order of magnitude.
This problem was recently revisited in Ref. \cite{gao97}, where
a comparative study was carried out for both the on-shell
properties and dilepton production rates using  
almost all the existing models for the $\pi\rho a_1$ dynamics.
The results were indeed surprising: although most models
provide reasonable description of the on-shell properties, the
corresponding dilepton rate could differ by two-and-half orders
of magnitude. By using the experimentally-constrained spectral
function \cite{huang95}, it was found that the effective
chiral Lagrangian of Ref. \cite{comm84}, in which the vector mesons
are introduced as massive Yang-Mills fields of the chiral
symmetry, provides the best off-shell, as well as on-shell,
properties of the $\pi\rho a_1$ dynamics. The cross sections
for $\pi a_1\rightarrow l{\bar l}$ from this model will 
be used in this work. 

The contributions from the secondary processes outlined above
are shown in the middel panel of Fig. 4. These are 
obtained in the relativistic transport model of Refs. \cite{likob,lib97},
including the HELIOS-3 acceptances, mass resolution, and
normalization \cite{helios}. It is seen that the $\pi a_1$
process is by far the most important source for dimuon
yields in this mass region. The $\pi\omega$ process also plays
some role in the entire intermediate-mass region. The contributions 
from $\pi\pi$, $\pi\rho$ and $K{\bar K}$ are important around 1 GeV
invariant mass.

In the right panel of Fig. 4, we add the contributions from the 
secondary processes obtained in our transport model to 
the background, and compare again with the HELIOS-3 data 
for central S+W collisions. It is seen that the data 
can now be nicely explained. Thus we showed for the first 
time the importance of the secondary processes
for the intermediate-mass dilepton spectra in heavy-ion 
collisions. This is an important step forward in the
use of intermediate-mass dilepton spectra as a probe of
the phase transition and QGP formation. Although the current
data do not show any necessity to invoke the QGP formation
in S-induced reactions, consistent with conclusions 
from the $J/\Psi$ physics, the observation that the secondary
processes do play an important role in the intermediate-mass
dilepton spectra is significant. 

\begin{figure}[hbt]
\begin{center}
\epsfig{file=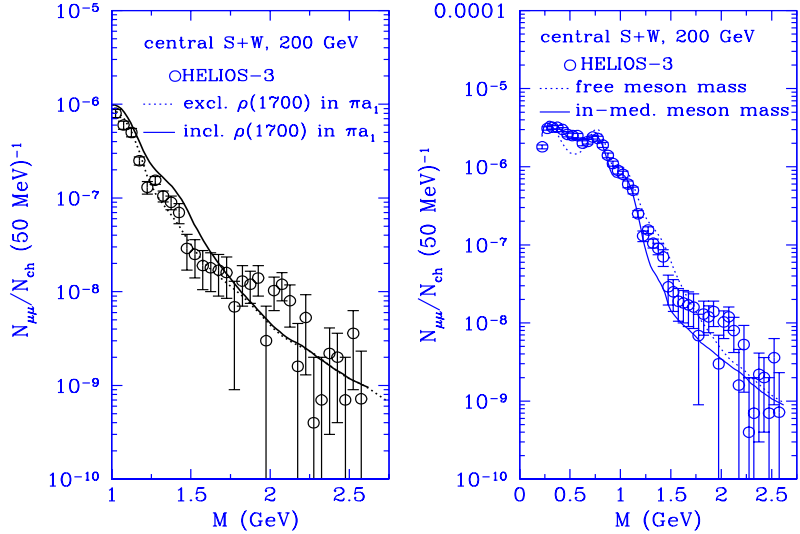,height=5in,width=3.5in,angle=270}
\caption{Left panel: comparison of dimuon spectra obtained with
the full form factor and with the $\rho (700)$ only.
Right panel: comparsion of dimuon spectra obtained 
with bare and in-medium meson masses, respectively.}
\end{center}
\end{figure}

It is appropriate to discuss the issue of the electromagnetic
form factor. While the form factors for $\pi\pi$, $\pi\rho$, $\pi\omega$,
$K{\bar K}$ and $K{\bar K^*}+c.c.$ are well determined
from experimental data for $e^+e^-$ annihilation, the
determination of the $\pi a_1$ electromanetic form factor
from the experimental data on $e^+e^-\rightarrow \pi^+\pi^-\pi^+\pi^-$
involves some uncertainties. In the previous calculation
we assumed that all the observed 4$\pi$ final state cross section
proceeds through the $\pi a_1$ intermediate state. There are
some indication that some of the cross section is dominated
by the $\rho (1700)$ which decays directly to $\rho\pi\pi$
without going through the $\pi a_1$ intermediate state \cite{pdata}. 
To see the sensitivity of our results on dimuon spectra to this
uncertainty, we also did a calculation in which the 
$\pi a_1$ form factor contains only the normal $\rho (770)$. The
cross section for $e^+e^-\rightarrow \pi a_1$ is then very
similar to that obtained in Ref. \cite{pen80}.
The results are shown in the left panel of Fig. 5 
by the dotted curve. 
Apparently, with a form factor that excludes the $\rho (1700)$
resonance, the contribution from the $\pi a_1$ process
is reduced. The agreement with the HELIOS-3 data in this
case is slightly better. 

Another issue we want to address here is the effects of dropping vector
meson masses on the entire dimuon spectra from the threshold to 
about 2.5 GeV. In the previous section we have shown that
the enhancement of low-mass dileptons indicates that the
vector meson masses decrease with increase density and temperature.
This will also affect the dilepton spectra in the intermediate-mass
region, through mainly two effects. One is the change of the
invariant energy spectra of these secondary meson pairs. If the rho meson
mass is reduced, then the invariant energy of the $\pi\rho$ 
collisions should also decrease. The second effect enters through
the modification of the electromagnetic form factor.
Since we do know very well how the masses of the $\rho$-like
resonances change with density and temperature, we assume
that they experience the same amount of the scalar field as
the ordinary rho meson, namely, $m_{\rho ,\rho^\prime} ^*
=m_{\rho ,\rho^\prime} -2/3g_\sigma\langle \sigma \rangle$. 
The results of this
calculation are shown in the right panel of Fig. 5.
Below 1.1 GeV and especially from 0.4 to 0.6 GeV, the agreement 
with the experimental data is much better when the dropping 
vector meson mass scenario is introduced. At higher masses, the
dropping mass scenarios somewhat underestimates the
experimental data. In this mass region, however, there might
be additional contributions from, e.g., secondary Drell-Yan 
processes \cite{spie97} that were not included in this study.
Also, the issues of collisional brodening and baryon contribution
need to be checked, at these invariant masses. These are
under study and will be reported on later.

\section{Direct photon: lack of the signal}

Single photon spectra in heavy-ion collisions at CERN-SPS energies
have been measured by the WA80 collaboration \cite{wa80}. So
far, only the upper bound has been determined for the so-called
`thermal' photon spectra, which accounts for about 5\% of the
total observed single photon yield, which is dominated by neutral
pion decay. In hydrodynamical calculations \cite{sinha94}, the absence
of significant thermal photons has been interpreted as an
evidence for the formation of a quark gluon plasma. Without
phase transition, the initial temperature of the hadronic gas
found in Ref. \cite{sinha94} is about 400 MeV. This leads to
a large number of thermal photons from hadronic interaction which
is not observed experimentally. Including the phase transition,
the initial temperature is lowered to about 200 MeV \cite{sinha94},
because of increased degrees of freedom. With a lower initial
temperature, the thermal photon yield is thus reduced and is found
to better agree with the WA80 data. However, a more recent analysis 
\cite{cley97} has shown that if one includes all hadron resonances with 
masses below 2.5 GeV the initial temperature can also be lowered to 
about 200 MeV, and the WA80 photon data can then be explained without 
invoking the formation of a quark gluon plasma. This is basically also the
conclusion of the detailed transport model analysis presented here.

\begin{figure}[htb]
\begin{center}
\epsfig{file=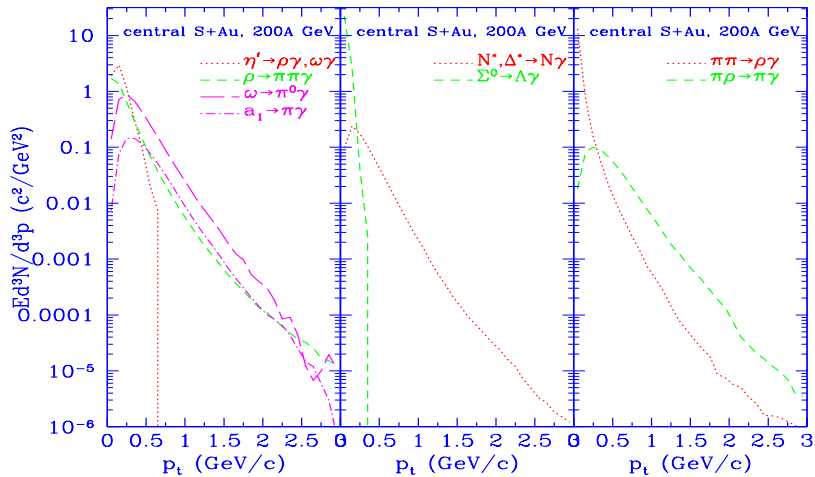,height=5in,width=3in,angle=270}
\caption{`Thermal' photon spectra in central S+Au collisions at 
200 AGeV from different sources.}
\end{center}
\end{figure}

For the `thermal' photon spectra we include the decay of $\rho$, $\omega$,
$\eta^\prime$, and $a_1$ mesons, as well as two-body processes such
as $\pi\pi\rightarrow \rho \gamma$ and $\pi\rho\rightarrow \pi\gamma$.
The decay width for $\rho\rightarrow \pi\pi\gamma$ is taken from the model
of Ref. \cite{sig63}, which describes well the measured width for 
$\rho^0\rightarrow \pi\pi\gamma$. The $\omega$ and $a_1$ radiative 
decay widths are proportional to $|{\bf p}_\pi |$, with the
coefficients determined from the measured width. For the two-body
cross sections, we use the results of \cite{kap91}, which do not
include the contribution from an intermediate $a_1$ meson. 
The latter has already been included in our model as a two-step process. 

The contributions to the so-called `thermal' photons
from the decay of mesons and baryons, and from two-body
scattering are shown in Fig. 6. The $\omega$ radiative
decay is found to be the most important source for photons
with transverse momenta above about 0.5 GeV. 
The contribution from the $a_1$ radiative decay is
somewhat larger than that from direct $\pi\rho\rightarrow \pi\gamma$.
This is in agreement with the conclusion of Ref. \cite{song93}
that including $a_1$ the $\pi\rho$ contribution increases 
by about a factor 2-3 in the relevant temperature and photon
energy region.

The contributions from $\eta^\prime$ and $\Sigma^0$
radiative decays are restricted to photons with transverse 
momenta below 0.5 GeV, because the mass differences between hadrons
in the intial and final states in these processes are small.
Photons with transverse momenta below 0.2 GeV come chiefly
from the decay of $\Sigma ^0$ and $\pi\pi$ scattering.
The reaction $\pi\pi\rightarrow \rho\gamma$ is endothermic,
with most of the available energy going into the rho meson mass.
This cross section actually diverges as the photon energy
goes to zero. We have included a low-energy cut-off as
in Ref. \cite{kap91}. The choice of this cut-off parameter
affects basically only the photons with transverse momenta
below about 0.1 GeV, where no experimental data are available.

\begin{figure}[htb]
\begin{center}
\vskip -1cm
\epsfig{file=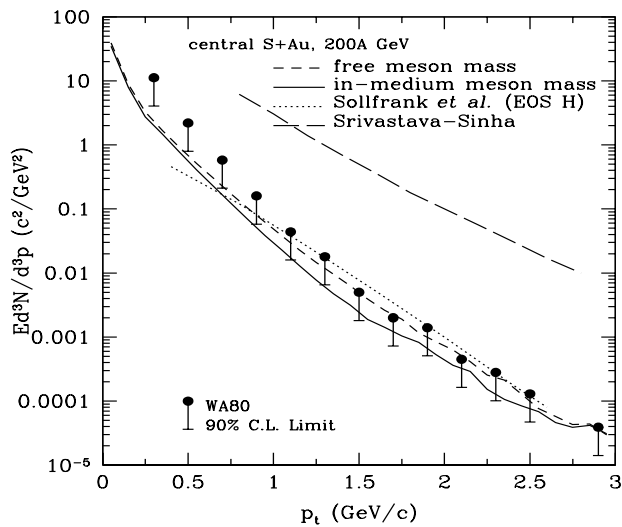,height=3.5in,width=4.0in}
\caption{`Comparison of `Thermal' photon spectra in central S+Au 
collisions at 200 AGeV from different calculations.}
\end{center}
\end{figure}

The importance of the radiative decay contribution from 
primary omega mesons was not seen in
thermal rate calculations such as those of Ref. \cite{kap91}. 
Non-equilibrium effects and dynamical evolution of the collision
attribute to the difference between the thermal and transport
model calculations. As the colliding system expands, the two-body
contribution becomes less important as compared with that
of one-body, since the former is proportional to density squared.
It was found in Ref. \cite{lib97} that in the first
a few fm/c, when the system is the densest, the contribution 
from $\pi\rho$ scattering is indeed more important than that 
from $\omega$ decay, in agreement with the findings of the thermal 
rate calculation in Ref. \cite{kap91}. However, the rate from 
$\pi\rho$ damps with time very fast. After 3-5 fm/c, the emission 
rate from $\omega$ meson becomes more important. Adding on top of 
this the contribution from the omega meson decay at the freeze out, 
it is easy to understand why the $\omega$ meson radiative decay 
becomes as important as that from $\pi\rho$ collisions in 
a dynamical calculation.

The WA80 single photon data have been looked at by several
groups, mostly based on hydrodynamical models 
In Fig. 7, we show our results together with those
from \cite{soll96} and \cite{sinha94}. In Ref. \cite{soll96}
several different equations of state with and without quark gluon
plasma formation were considered. Shown in the figure by dotted line
is their results based on EOS H. This equation of state
included about the same number of hadron degrees of freedom as
in our transport model. It is very interesting to see that
their results are very similar to ours, although the dynamical
models used are quite different. In Ref. \cite{sinha94}
two different equations of state, one with quark gluon
plasma formation and one with pure pionic gas, were
considered. Shown in the figure is their results 
based on the pionic gas equation of state. Because of
very limited number of hadronic degrees of freedom in
this equation of state, they needed a very high 
initial temperature to account for the final observed
hadron abundances. This led to large photon yield.

\section{SUMMARY AND OUTLOOK}
 
In summary, we have studied dilepton and photon production from both 
proton-nucleus and nucleus-nucleus collisions using the 
relativistic transport model with initial conditions 
determined by string fragmentation from the initial stage 
of the RQMD model.  It is found that the dilepton spectra 
in proton-nucleus reactions measured by the CERES and the HELIOS-3 
collaboration can be well understood in terms of conventional 
mechanisms of Dalitz decay, direct vector meson decay, decay of
charmed mesons, and initial Drell-Yan processes. 
For dilepton spectra in central heavy-ion collisions, these 
conventional mechanisms, however, fail to explain the data, 
in the entire mass region from threshold to about 2.5 GeV.

Including the contribution from pion-pion 
annihilation, which is important in the mass region from 
$2m_\pi$ to $m_{\rho ,\omega}$, removes some of the discrepancy.  
But the theoretical prediction is still substantially below the 
data in the low mass region and somewhat above the data around 
$m_{\rho,\omega}$. The theoretical results are brought into good 
agreement with the data when reduced in-medium vector meson masses 
are taken into account. In the intermediate-mass region, 
we have included additional secondary processes involving 
meson-meson collisions. The cross sections for these processes 
are constrained by the experimental data for $e^+e^-$ annihilation. 
We found that these secondary processes, especially the 
$\pi a_1 \rightarrow l{\bar l}$, play very important role 
in the intermediate-mass region. We have also calculated, within 
the same dynamical model, the thermal photon spectra in 
central S+Au collisions. In both free meson masses and 
in-medium meson masses scenarios, our results do not exceed 
the upper limit deduced from the experiments by the WA80 collaboration. 
 
So far, we have not included explicitly collisional brodening 
in our calculation. Since the rho meson is treated as a dynamical 
particle in our transport model, its collisions with meson 
(mainly pions) and baryons (mainly nucleons) are included. 
This might reflect, at least partly, its collisional brodening 
in hot and dense matter. On the other higher resonances, such as 
$\rho (1450)$ and $\rho (1700)$, are not treated dynamically.
Their effects are included through electromagnetic form factors. 
Therefore, collisional brodening is not considered for these 
higher resonances. Since they usually have a natural width of
200-300 MeV, a collsional brodening width of about 50 MeV
for these higher resonances \cite{hag95} shall not affect our 
results dramatically. This, however, needs to be dealt with 
quantitatively.

The current investigation can be extended to higher incident energies,
such as those of RHIC collider, by combining the cross sections
(or thermal rates) obtained in this study with, e.g., 
hydrodynamical models for the evolution of heavy-ion collisions
at the RHIC energies. This kind study will be very useful for the
determination of hadronic background in the dilepton 
spectra, and for the clear identification of dilepton
spetra from the QGP.   

\section*{Acknowledgments}
This work is supported in part by the U.S. Department of Energy
under grant number DE-FG02-88ER-40388,  by the Natural Sciences
and Engineering Research Council of Canada, and by the Fonds
FCAR of the Quebec Goverment.


\section*{References}

\begin{thebibliography}{9}

\bibitem{qm96} For example, Quark Matter'97 {\it Nucl. Phys}.
A590 (1995); Quark Matter'96, {\it Nucl. Phys}. A610 (1996).

\bibitem{shur80} E. Shuryak, {\it Phys. Rep}. 67 (1980) 71.

\bibitem{kkmm86} K. Kajantie, J. Kapusta, L. McLerran, and
A. Mekjian, {\it Phys. Rev. D} 34 (1986) 2746.

\bibitem{brown96} G.E. Brown and M. Rho, {\it Phys. Rep}. 269 (1996) 333.

\bibitem{ceres} G. Agakichiev {\it et al.}, {\it Phys. Rev. Lett}. 75
(1995) 1272; J.P. Wurm for the CERES Collaboration, {\it Nucl. Phys}. 
A590 (1995) 103c; I. Tserruya, {\it Nucl. Phys}. A590
(1995) 127c; G. Agakichiev {\it et al.,} {\it Nucl. Phys}. 
A610 (1996) 317c. 

\bibitem{drees96} A. Drees, {\it Nucl. Phys}. A610 (1996) 536c;
{\it Nucl. Phys}. (1998), in press. 

\bibitem{helios} M. Masera for the HELIOS-3 Collaboration, {\it Nucl. 
Phys}. A590 (1995) 93c; A.L.S. Angelis, {\it et al} CERN-PPE/97-117. 

\bibitem{na38} M.C. Abreu {\it et al.,} (NA38 collaboration), 
{\it Phys. Lett. B} 368 (1996) 230; 
M.C. Abreu {\it et al.,} (NA50 collaboration), {\it Nucl. Phys}. 
A610 (1996) 331c; C. Lourenco, {\it Nucl. Phys}. A610 (1996) 552c.

\bibitem{likob} G.Q. Li, C.M. Ko, and G.E. Brown, {\it Phys. Rev. Lett}.
75 (1995) 4007; {\it Nucl. Phys}. A606 (1996) 568; G.Q. Li, C.M. Ko,
G.E. Brown, and H. Sorge, {\it ibid.} A611 (1996) 539.

\bibitem{cassing} W. Cassing, W. Ehehalt, and C. M. Ko, {\it Phys. Lett. 
B} 363 (1995) 35;  W. Cassing, W. Ehehalt, and I. Kralik, {\it Phys. Lett .B}
377 (1996) 5.

\bibitem{rapp}  G. Chanfray, R. Rapp, and J. Wambach, {\it Phys. Rev. Lett.}
76 (1996) 368; R. Rapp, G. Chanfray, and J. Wambach, {\it Nucl. Phys}. 
A617, 472 (1997); 

\bibitem{sri96} D. K. Srivastava, B. Sinha, and C. Gale, 
{\it Phys. Rev. C} 53 (1996) R567. 

\bibitem{soll96} J. Sollfrank, {\it et al.,} {\it Phys. Rev. C} 
55 (1997) 392.

\bibitem{hung97} C. M. Hung and E. Shuryak, {\it Phys. Rev. C} 56 
(1997) 453.

\bibitem{sch96} H.J. Schulze and D. Blaschke, {\it Phys. Lett. 
B} 386 (1996) 429.

\bibitem{other} K. Haglin, {\it Phys. Rev. C} 53 (1996) R2606;
V. Koch and C. S. Song, {\it Phys. Rev. C} 54 (1996) 1903;
J.V. Steele, H. Yamagishi, and I. Zahed, {\it Phys. Lett. B} 384 (1996) 255;
R. Baier, M. Dirks, and K. Redlich, {\it Phys. Rev. D} 55 (1997) 4344.

\bibitem{br91} G.E. Brown and M. Rho, {\it Phys. Rev. Lett.}
66 (1991) 2720.

\bibitem{hat92} T. Hatsuda and S.-H. Lee, {\it Phys. Rev. C} 46 (1992) R34.

\bibitem{satz86} T. Matsui and H. Satz, {\it Phys. Lett. B} 178
(1986) 416.

\bibitem{na50} M.C. Abreu {\it et al.,} (NA50 collaboration), 
{\it Nucl. Phys.} A610 (1996) 404c; D. Kharzeev, 
{\it Nucl. Phys}. A610 (1996) 418c;
C.Y. Wong, {\it Nucl. Phys.} A610  (1996) 434c;
S. Gavin and R. Vogt, {\it Nucl. Phys.} A610  (1996) 442c;
J.-P. Blaizot and J.-Y. Ollitrault, {\it Nucl. Phys.} A610
(1996) 452c.

\bibitem{shur78} E. Shuryak, {\it Phys. Lett. B} 79 (1978) 135.

\bibitem{kap91} J. Kapusta, P. Lichard, and D. Seibert, 
{\it Phys. Rev. D} 44 (1991) 2774.

\bibitem{wa80} R. Albrecht {\it et al.,} {\it Phys. Rev. Lett.}
76 (1996) 3506.

\bibitem{baur96} R. Baur {\it et al.,} {\it Z. Phys. C} 71 (1996) 577.

\bibitem{sinha94} D. K. Srivastava and B. Sinha, {\it Phys. Rev. Lett.} 
73 (1994) 2421. 

\bibitem{dumi95} A. Dumitru {\it et al.,} {\it Phys. Rev. C} 
51 (1995) 2166.

\bibitem{cley97} J. Cleymans, K. Redlich, and D.K. Srivastava,
{\it Phys. Rev. C} 55 (1997) 1431.

\bibitem{lib97} G.Q. Li and G.E. Brown, Nucl. Phys. A, submitted;
M.A. Halasz, J.V. Steele, G.Q. Li, and G.E. Brown, nucl-th/9712006.

\bibitem{koli96} C.M. Ko and G.Q. Li, {\it J. Phys. G} 22 (1996) 1673.

\bibitem{qhd86} B.D. Serot and J.D. Walecka, {\it Adv. Nucl. Phys.}
16 (1986) 1. 

\bibitem{sorge89} H. Sorge, H. St\"ocker, and W. Greiner, {\it Ann. Phys.} 
192 (1989) 266.

\bibitem{thomas94} K. Saito and A.W. Thomas, {\it Phys. Rev. C} 51
(1995) 2757.
 
\bibitem{land85} L. G. Landberg, {\it Phys. Rep.} 128 (1985) 301.

\bibitem{rqmddi} M. Hoffmann {\it et al.,} {\it Nucl. Phys.} 
A566 (1994) 15c.

\bibitem{liko95} G.Q. Li and C.M. Ko, {\it Nucl. Phys.} 
A582 (1995) 731.

\bibitem{braun97} P. Braun-Munzinger, D. Miskowiec, A. Drees, and C. Lourenco,
CERN-PPE/97-65, {\it Z. Phys. C}, in press.

\bibitem{gale94} C. Gale and P. Lichard, {\it Phys. Rev. D} 49
(1994) 3338.

\bibitem{song94} C. Song, C.M. Ko, and C. Gale, {\it Phys. Rev. D} 
50 (1994) R1827.

\bibitem{haglin95} K. Haglin and C. Gale, {\it Phys. Rev. D} 52
(1995) 6297.

\bibitem{kim96} J.K. Kim, P. Ko, K.Y. Lee, and S. Rudaz,
{\it Phys. Rev. D} 53 (1996) 4787.

\bibitem{xiong92} L. Xiong, E.V. Shuryak, and G.E. Brown, 
{\it Phys. Rev. D} 46 (1992) 3798.

\bibitem{song93} C. Song, {\it Phys. Rev. C} 47 (1994) 3338.

\bibitem{bia91} M.E. Biagini, S. Dubnicka, E. Etin, and P. Kolar,
{\it Nuovo Cimento A} 104 (1991) 363.

\bibitem{donn89} A. Donnachie and A.B. Clegg, {\it Z. Phys. C} 42 (1989)
663.

\bibitem{mane82} F. Mane {\it et al.,} {\it Phys. Lett. B} 112 (1982) 
179.

\bibitem{doli91} S.I. Dolinsky {\it et al.,} {\it Phys. Rep.} 202 (1991) 99.

\bibitem{gao97} C. Gale, nucl-th/9706026;
S. Gao and C. Gale, {\it Phys. Rev. C}, in press.

\bibitem{huang95} Z. Huang, {\it Phys. Lett. B} 361 (1995) 131. 

\bibitem{comm84} H. Gomm, O. Kaymakcalan, and J. Schechyter,
{\it Phys. Rev. D} 30 (1984) 2345.

\bibitem{pdata} Particle Data Group, Review of Particle Properties,
{\it Phys. Rev. D} 50 (1994) 1173.

\bibitem{pen80} G. Penso and T.N. Truong, {\it Phys. Lett.} 95B (1980)
143.

\bibitem{spie97} C. Spieles {\it et al.,} to be published.

\bibitem{sig63} P. Singer, {\it Phys. Rev.} 130 (1963) 2441.

\bibitem{hag95} K. Haglin, {\it Nucl. Phys.} A584 (1995) 719.
 
\end{thebibliography}
\end{document}